\documentclass[aps,pre,twocolumn,10pt,amsmath,amsfonts,amssymb,floatfix,bibnotes]{revtex4-1}
\usepackage{amssymb}
\usepackage{graphicx,color}
\usepackage{multirow}

\usepackage{amsmath,bm,epsfig}
\newcommand{\B}[1]{{\bm{#1}}}
\newcommand{\C}[1]{{\mathcal{#1}}}    

\newcommand{\beq}{\begin{equation}}
\newcommand{\eeq}{\end{equation}}
\newcommand{\ba}{\begin{eqnarray}}
\newcommand{\ea}{\end{eqnarray}}

\begin{document}

\title{Scaling Theory of Giant Frictional Slips in Decompressed Granular Media}

\author{H. George E. Hentschel$^{1.2}$, Itamar Procaccia$^1$ and Saikat Roy$^1$}
\affiliation{$^1$Dept. of Chemical Physics, the Weizmann Institute of Science, Rehovot 76100, Israel\\$^2$Dept. of Physics, Emory University, Atlanta Georgia. }

\begin{abstract}
When compressed frictional granular media are decompressed, generically a fragile
configuration is created at low pressures. Typically this is accompanied by
a giant frictional slippage as the fragile state collapses. We show that this
instability is understood in terms of a scaling theory with theoretically computable
amplitudes and exponents. The amplitude diverges in the thermodynamic limit hinting to the possibility of huge frictional slip events in decompressed granular media. The physics of this slippage is discussed in terms of the probability distribution functions of the tangential and normal forces on the grains which are highly correlated due to the Coulomb condition.
\end{abstract}

\maketitle

The physics of frictional granular matter are not abundant with universal results. Under the action of external stresses, particles in frictional granular matter form an inhomogeneous contact network, which supports the external load by creating force chains \cite{98CWBC}. The nature of the network of force chains depends on the type of external load, resulting in a different geometry for shearing and for isotropic compression \cite{05MB}. The presence of friction complicates matters, excluding any Hamiltonian description and introducing history dependence for the unavoidable dissipative events that occur under any type of loading. Further stressing results in modifying the force chains that typically break and re-form while energy is dissipated. Upon compression and decompression hysteresis is typical, and elasticity theory alone cannot provide an adequate description of the complex physics of these materials \cite{08CCGP,11KN,14PKN,14KA,13BRKE,13RLR,14FFS,18BHPRZ}. Thus discovering universal aspects within this complexity is important, promising some stable hinges around which a theory of the mechanical properties of frictional granular matter may be erected. The purpose of this Letter is to reveal one such universal phenomenon which has to do with the number of frictional slip events that occur upon decompression; This number diverges (in the
thermodynamic limit) when the pressure goes to zero as a power law with an amplitude
and an exponent that are both theoretically computable.

\begin{figure}
\includegraphics[scale=0.20]{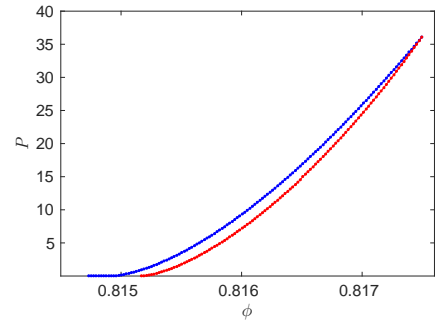}
\caption{A typical hysteresis loop showing the pressure as a function of the
packing fraction $\phi$. Here $N=4000$ and $P_{\rm max}=36.1$. The compression leg is
shown in blue and the decompression leg in red. Below we focus
on the frictional slip events that occur along the decompression leg. }
\label{loop}
\end{figure}

\begin{figure}
\includegraphics[scale=0.20]{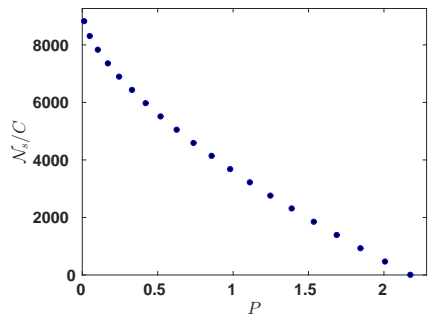}
\caption{The cumulative number of frictional slips ${\cal N}_s (N,P_{\rm max},P)/C$ as a function
of $P$ averaged over 10 independent decompression legs. The maximal pressure $P_{\rm max}$ averaged
 over these 10 legs is $P_{\rm max}=2.18$ and $N=4000$. Note the apparent
divergence of the cumulative number when the pressure reduces to zero.}
\label{result}
\end{figure}

To focus the discussion consider a typical compression-decompression hysteresis loop that is
obtained by uniaxially compressing an array of $N$ bi-dispersed disks of unit mass with diameters $a_1=1$ and $a_2=1.4$ respectively. All lengths below will be measured in units of $a=a_1$. The particles are placed randomly in a three dimensional box of dimension, $57$ (along $x$), $102$ (along $y$) and $1.4$ (along $z$). Quasistatic compression and decompression are implemented by displacing the boundary particles. A side wall made of particles is placed in the direction perpendicular to the compression direction. The normal force acting on the contact between disks $i$ and $j$,  $\B F^{(n)}_{ij}$, is modeled as a Hertzian contact, whereas the tangential force $\B F^{(t)}_{ij}$ is given by a Mindlin force \cite{79CS}. Keep in mind that both the magnitudes of the normal forces $F^{(n)}_{ij}$ and the tangential forces $F^{(t)}_{ij}$ are random variables, but at each contact they must obey  the Coulomb law $F^{(t)}_{ij}\le \mu F^{(n)}_{ij}$ where $\mu$ is the friction coefficient for mechanical stability to exist. Finally also note that the Hertzian force is derivable from
 a potential but the Mindlin force is not. The scale of the normal potential will be denoted by
 $\epsilon$ and all the energy scales will be measured in units for which $\epsilon=1$.  In Fig.~\ref{loop} we see a typical hysteresis loop that is obtained by compressing the system from pressure $P=0$ to a maximal value of the pressure $P_{\rm max}$ and then decompressing back to zero pressure. The pressure is plotted as a function of the packing fraction $\phi$ which is the ratio
of the total disk area to the area of the containing box.
Upon decompression the normal forces reduce, and consequently the Coulomb law can be violated at a contact and a frictional slip event then takes place to restore this law.  Details of the simulation technique are available in the methods section below, but here it is important to state that in the present simulations we used the convention \cite{07SHS} that the frictional force remains at the Coulomb limit when a slip occurs unless the normal force is reduced again, and see more about this below. We focus now on the decompression leg and ask how many frictional slip events ${\cal N}_s$ occur while we decompress from any chosen pressure value $P_{\rm max}$ to any other given pressure $P$. In other words, we measure
\begin{equation}
{\cal N}_s (N,P_{\rm max},P) \equiv \int_P^{P_{\rm max}} n(N,P) dP \,
\label{defNs}
\end{equation}
where $n(N,P) dP$ are the number of frictional slips that occur when decompressing
from $P+dP$ to $P$:
\begin{equation}
\label{defn}
n(N,P) \equiv -\frac{d{\cal N}_s (N,P_{\rm max},P)}{dP} \ .
\end{equation}
An example of the result of the measurement of ${\cal N}_s (N,P_{\rm max},P)/C$ as a function of $P$ is shown in Fig.~\ref{result}. The simulation indicates an apparent divergence of the cumulative
number of slip events as the pressure reduces towards zero. Our aim here is to understand and
quantify this finding.

To understand the phenomenon we realize that as $P\to 0$ the system is approaching
the ``unjamming" transition where the packing fraction $\phi$ approaches the jamming
packing fraction, $\phi\to \phi_J(\mu)$ \cite{07SHESS}. In frictional media the jamming packing
fraction depends of course on the friction coefficient;  nevertheless in its vicinity there exists a divergent rigidity length scale \cite{Wyart}
\begin{equation}
\xi(P)/a \sim (P/P_0)^{-\zeta} \ ,
\label{xiP}
 \end{equation}
 where $P_0 \equiv \epsilon/a^d$ is the pressure scale set by the intergranular normal potential (and thus $P_0=1$ in units in which $a=1$ and $\epsilon =1$). The rigidity scales separates two different modes of behavior. For length $\ell$ such that $\ell<\xi(P)$
the amorphous matter contains unstable floppy modes;  for length scales
$\ell>\xi(P)$ the amorphous matter supports the external stress and can be treated using continuum elasticity theory. The central idea proposed below is that this very same
length scale also controls the frictional slippage. The novelty of this idea is in connecting
two seemingly disparate subjects, that of jamming in granular media and plasticity in these media
due to frictional dissipation.

To establish the connection between the rigidity length-scale and frictional slips we recall some known facts are about the former. As $\phi \rightarrow \phi_J(\mu )$ simulations show \cite{11VVMOT} that this length scale diverges as
\begin{equation}
\xi(P) \sim [\phi(P) - \phi_J(\mu )]^{-\nu} a \ .
\label{xiphi}
\end{equation}
 Simultaneously, the pressure goes to zero as
 \begin{equation}
 P \sim P_0 [\phi(P) - \phi_J(\mu )]^{\psi} \ ,
  \label{Pphi}
  \end{equation}
  where the exponent $\psi = \alpha -1$, where $\alpha$ is the exponent characterizing  the intergranular normal force as the grains are squeezed.  For a Hertzian force therefore we have  $\psi = 3/2$.  Combining Eqs.~(\ref{xiP})-(\ref{Pphi}) determines the exponent $\zeta$ in Eq.~(\ref{xiP}) as
\begin{equation}
\label{zeta}
\zeta = \nu/\psi  \ .
\end{equation}

\begin{figure}
\includegraphics[scale=0.40]{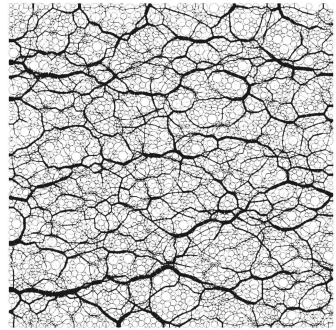}
\caption{Typical force chains in stressed frictional granular media}
\label{forcechains}
\end{figure}

To make the connection to frictional slips discussed above we argue that the same rigidity length scale also controls the number of slips events at pressure $P$. The modes responsible for the frictional slips are the unstable floppy modes that lead to localized avalanches of frictional slips.
In other words, at scales $l< \xi (P)$ there exist local slip avalanches everywhere in the granular medium. If we look at a typical sample of the stress network distribution (see Fig.~\ref{forcechains}), we see that the most likely places for slippage will occur is in the regions of the force chain network where the stress is a maximum. The fractal dimension $d_s$ of the backbone clearly obeys $1\le d_s\le d$ where $d$ is the system's dimension. In the two dimensional simulation ($d=2$) it appears that $d_s =1$.

We thus propose that slip events can be considered as avalanches of lengthscale $\xi(P)$ containing $(\xi(P)/a)^{d_s}$ grains, associated with the localized force network which becomes unstable upon decompression. Moreover, if the system is of size $L^d$ and it contains $N$ disks of average
diameter $a$ then $N\approx (L/a)^d$. The system contains $(L/\xi(P))^d$ regions each of which
contains $(\xi(P)/a)^{d_s}$ slipping disks. A consequence of this thinking is that $n(N,P)$ is expected to scale as
\begin{eqnarray}
\label{slip}
\!n(N,P) &\sim& \frac{1}{P}\! \left(\frac{L}{\xi(P)}\right)^d \left(\frac{\xi(P)}{a}\right)^{d_s}\!\!\! =  \! \frac{1}{P} \left(\frac{L}{a}\right)^d \!\left(\frac{\xi(p)}{a}\right)^{d_s-d}\nonumber \\
& \!\!\!\sim &\frac{1}{P} N \left(\frac{\xi(P)}{a}\right)^{d_s-d}\sim \frac{N}{P_0} \left(\frac{P}{P_0}\right)^{\zeta (d-d_s)-1} \ ,
\end{eqnarray}
where the $P^{-1}$ factor is due to the definition in Eq.~(\ref{defn}).
 Now substituting this expression into Eq.~(\ref{defNs}) and integrating we find
\begin{eqnarray}
\label{int2}
&&{\cal N}_s(N,P,P_{\rm max}) \sim N [(P_{\rm max}/P_0)^{\zeta(d-d_s)}-(P/P_0)^{\zeta(d-d_s)}]\nonumber\\ &&\approx C N (P_{\rm max}/P_0)^{\zeta(d-d_s)}[1-(P/P_{\rm max})^{\zeta(d-d_s)}].
\end{eqnarray}
where $C$ is a constant of the order of the number of decompression steps going from $P_{\rm max}$ to $P_0$. Note that C depends in numerics on the decompression step sizes $\Delta P$.
\begin{figure}
\includegraphics[scale=0.20]{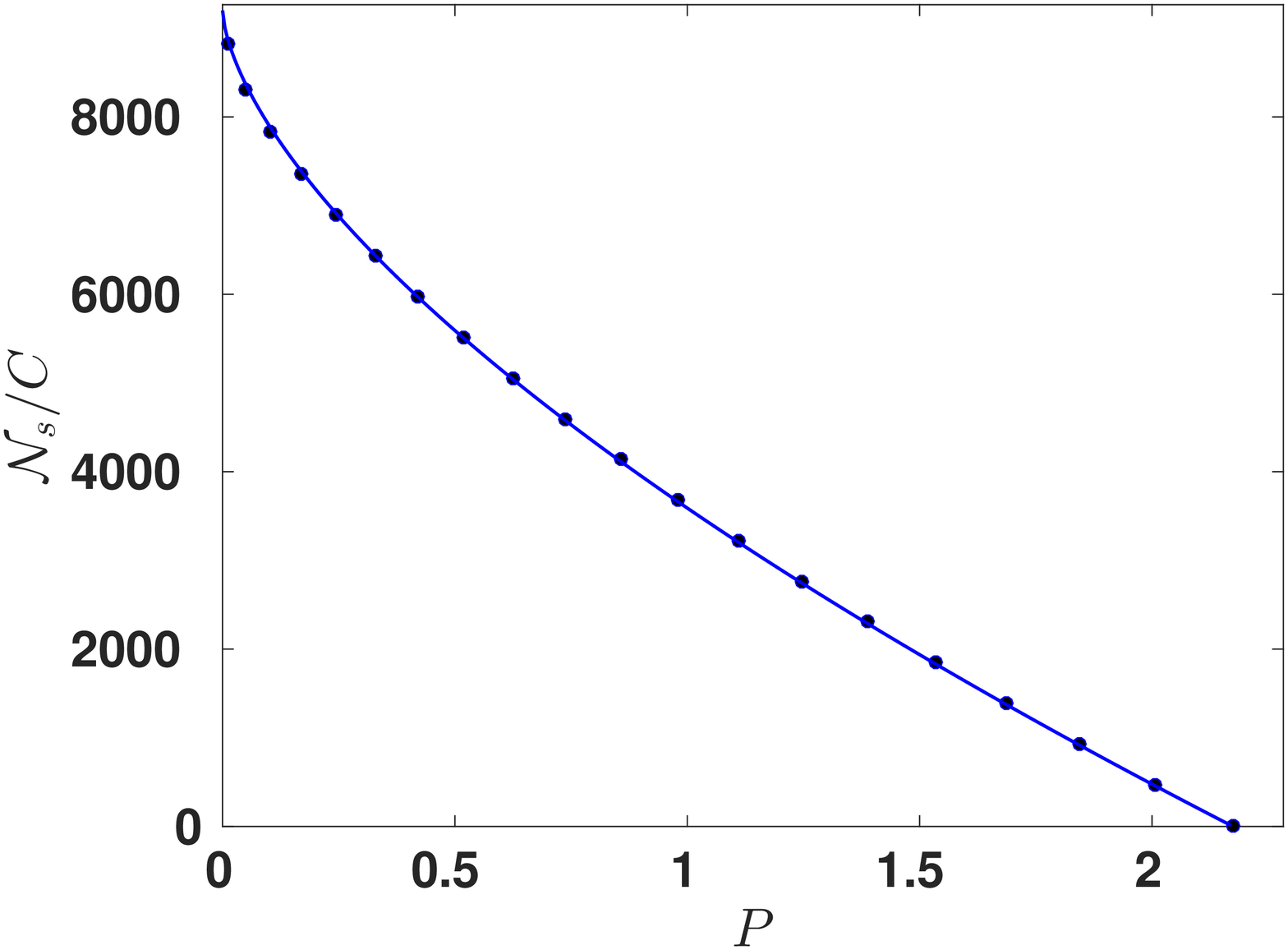}
\includegraphics[scale=0.20]{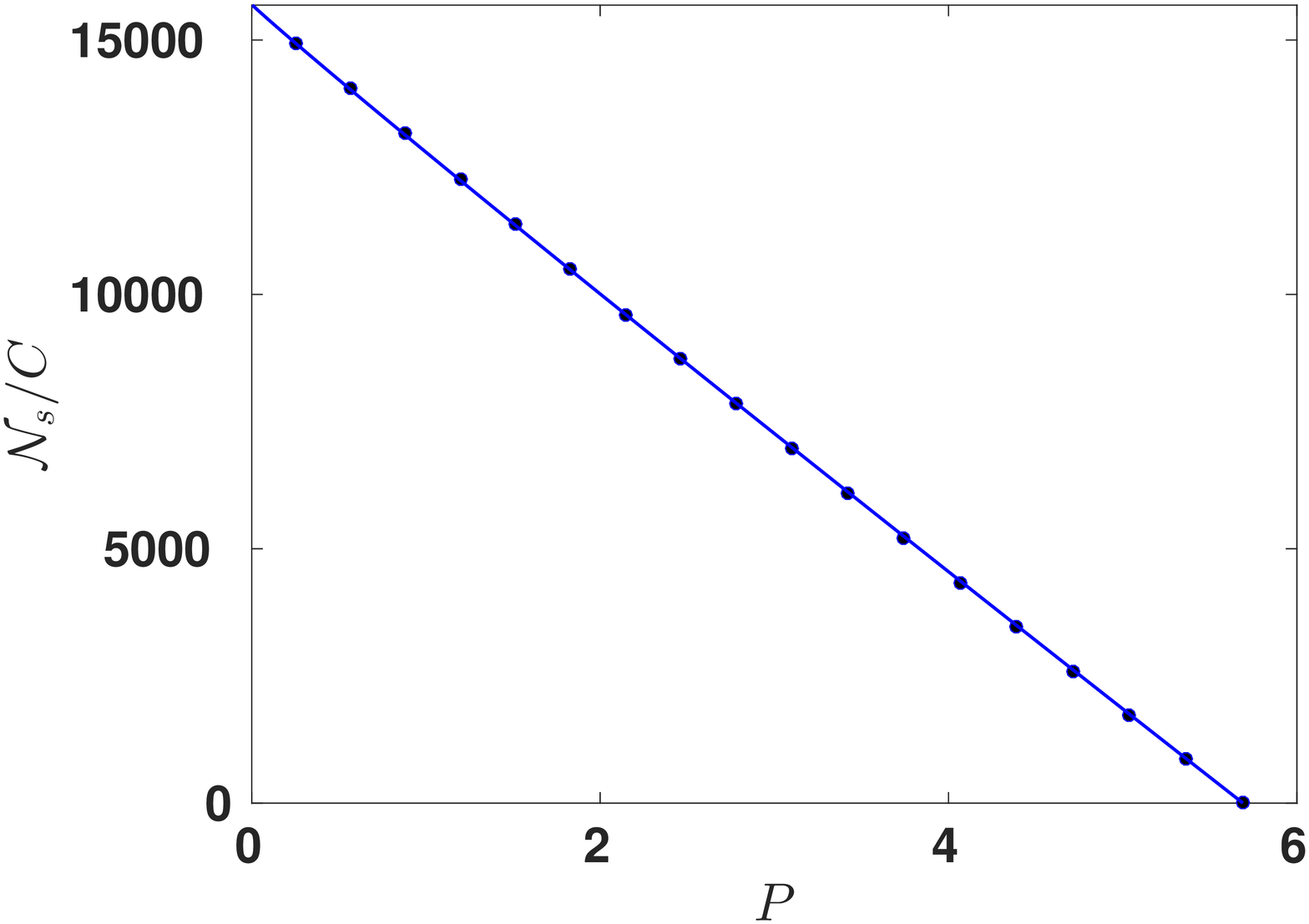}
\caption{Comparison of theory (continuous line) and simulations (points) for the cumulative number of frictional slips ${\cal N}_s (N,P_{\rm max},P)/C$ as a function
of $P$, fitted to the theoretical result Eq.~(\ref{form}) in the
range $P\le P_{\rm max}$. Upper panel: the Hertzian case of Fig.~\ref{result} with $x= 0.64$. Lower panel: Hookean case
with $x=0.97$ and $P_{\rm max}=5.7$. }
\label{test}
\end{figure}

To simplify the comparison of the theoretical result Eq.~(\ref{int2}) to the numerical
observation shown in Fig.~\ref{result} we rewire it in terms of an amplitude and an exponent:
\begin{equation}
\frac{{\cal N}_s(N,P,P_{\rm max})}{C}\approx  A(N,P_{\rm max}) [1-(P/P_{\rm max})^x] \ ,
\label{form}
\end{equation}
where the amplitude $A\approx N (P_{\rm max}/P_0)^x$ and $x=\zeta(d-d_s)=\nu(d-d_s)/\psi $.
For both Hertzian and Hookean contacts one expects \cite{11VVMOT} $\nu\approx 1$ and
following the discussion above for the two dimensional simulations we use $d_s=1$. Then $x\approx 0.66$ and the amplitude is
\begin{equation}
\label{res1}
A(N,P_{\rm max}) \approx N (P_{\rm max}/P_0)^x \ , \quad x \approx 0.66 \ ,
\end{equation}
up to constants of the order of unity.
A test of this result is presented in the upper panel of Fig.~\ref{test} in which Eq.~(\ref{form}) is fitted to the data shown in Fig.~\ref{result} in the region $P\le P_{\rm max}$. The continuous line is a two parameter least-squares fit in which $x=0.64$.
The fitted amplitude is $A(N,P_{\rm max})=9200$, of the order of the prediction of Eq.~(\ref{res1}) which is 6713. We conclude that
the agreement between the data and theory is quite satisfactory. One should stress that the
theory predicts a diverging amplitude $A(N,P_{\rm max})$ in the thermodynamic limit
$N\to \infty$. Whether this is related to the ``scale-free" nature of geological gigantic
earth quake events or not remains to be seen in future research.
\begin{figure}
\includegraphics[scale=0.20]{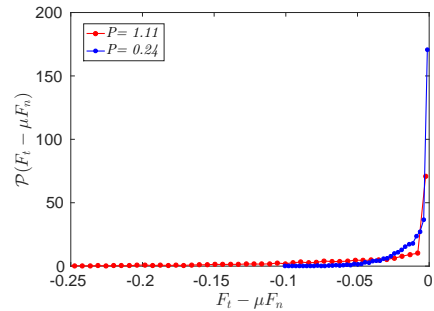}
\caption{The pdf $P(F_t-\mu F_n)$ for two different pressures during the decompression step.
The existence of the Coulomb constraints results in a growing divergence of the pdf near zero,
indicating the fragility of the configuration and the eminent giant slip events. Here $\mu = 0.1$.}
\label{pdf}
\end{figure}

To increase our confidence in the theory presented above we repeated the simulations with
Hookean instead of Hertzian contact normal forces. All that changes is the prediction for the
exponent $x$ which changes from $0.66$ to 1.0 because now $\psi=1$. The comparison of simulations
and theory is shown in the lower panel of Fig.~\ref{test}. The exponent of the best fit is $x=0.97$
$A(N,P_{\rm max})=15700$ in good agreement with Eq.~(\ref{res1}) which predicts for the amplitude the
value of 22762. In both cases (Hertzian and Hookean) coefficients of the order of unity in the
theoretical expectation Eq.~(\ref{res1}) are certainly acceptable.

The phenomenon described here is intimately connected to the unavoidable correlations between
normal and tangential forces in frictional granular matter due to the Coulomb condition. It was shown
recently \cite{18ABHPS} that a maximum entropy formalism for the joint probability distribution functions (pdf's) of the normal and tangential forces predicts that at decreasing pressures there will be a tendency for a divergence near zero of the pdf $\C P(F_t-\mu F_n)$. This pdf for our present case of Hertzian contacts for two different pressures is show in Fig.~\ref{pdf}. One can see the growing divergence near zero which is a clear indication of the frictional fragility of the configuration, with more and more contacts having residing near their Coulomb threshold. This is the fundamental physics
responsible for the giant slip event discussed in the Letter.

{\bf Methods}: For frictional granular matter Molecular Dynamics simulations are used to correctly keep track of both the normal and the (history dependent) tangential forces \cite{01SEGHLP,16GPP}. We have set up simulation of uniaxial compression of two dimensional granular packings, performed using open source codes, LAMMPS \cite{95P} and LIGGGHTS \cite{12KGHAP}. The particles are taken as bi-dispersed disks of unit mass with diameters 1 and $1.4$ respectively. All the lengths in the simulations are measured in units of the small diameter. The particles are placed randomly in a three dimensional box of dimension, $57$ (along $x$), $102$ (along $y$) and $1.4$ (along $z$). Quasistatic compression is implemented by displacing the boundary particles. A side wall made of particles is placed in the direction perpendicular to the compression direction.

The contact forces (both the normal and tangential forces which arise due to friction) are modeled according to the discrete element method developed by Cundall and Strack \cite{79CS}.
When the disks are compressed they interact via both
normal and tangential forces.  Particles $i$ and $j$, at positions ${\B r_i, \B r_j}$ with velocities ${\B v_i, \B v_j}$ and angular velocities ${\B \omega_i, \B \omega_j}$ will  experience a relative normal compression on contact given by $\Delta_{ij}=|\B r_{ij}-D_{ij}|$, where $\B r_{ij}$ is the vector joining the centers of mass and $D_{ij}=R_i+R_j$; this gives rise to a  normal force $ \B F^{(n)}_{ij} $. The normal force is modeled as a Hertzian contact, whereas the tangential force is given by a Mindlin force \cite{79CS}. Defining $R_{ij}^{-1}\equiv R_i^{-1}+R_j^{-1}$, the force magnitudes are,
\begin{eqnarray}
\B F^{(n)}_{ij}&=&k_n\Delta_{ij} \B n_{ij}-\frac{\gamma_n}{2} \B {v}_{n_{ij}}\ , \:
\B F^{(t)}_{ij}=-k_t \B t_{ij}-\frac{\gamma_t}{2} \B {v}_{t_{ij}} \\
k_n &=& k_n^{'}\sqrt{ \Delta_{ij} R_{ij}} \ , \quad
k_t = k_t^{'} \sqrt{ \Delta_{ij} R_{ij}} \\
\gamma_{n} &=& \gamma_{n}^{'}  \sqrt{ \Delta_{ij} R_{ij}}\ , \quad
\gamma_{t} = \gamma_{t}^{'}  \sqrt{ \Delta_{ij} R_{ij}} \ .
\end{eqnarray}
Here $\delta _{ij}$ and $t_{ij}$ are normal and tangential displacement; $R_{ij}$  is the effective radius. $\B n_{ij}$ is the normal unit vector.  $k_n^{'}$ and $k_t^{'}$ are spring stiffness for normal and tangential mode of deformation: $\gamma_n^{'}$ and $\gamma_t^{'}$ are viscoelastic damping constant for normal and tangential deformation. We also used a Hookean model where now $ k_n = k_n^{'}, k_t = k_t^{'} , \gamma_{n} = \gamma_{n}^{'} , \gamma_{t} = \gamma_{t}^{'} $ .
   $\B {v_n}_{ij}$ and $\B {v_t}_{ij}$ are respectively normal and tangential component of the relative velocity between two particles. The relative normal and tangential velocity are given by
   \begin{eqnarray}
\B {v}_{n_{ij}}&=& (\B {v}_{ij} .\B n_{ij})\B n_{ij}  \\
\B {v}_{t_{ij}}&=& \B {v}_{ij}-\B {v}_{n_{ij}} - \frac{1}{2}(\B \omega_i + \B \omega_j)\times \B r_{ij}.
\end{eqnarray}
   where $\B {v}_{ij} = \B {v}_{i} - \B {v}_{j}$. Elastic tangential displacement $ \B t_{ij}$ is set to zero when the contact is first made and is calculated using $\frac{d \B t_{ij}}{d t}= \B {v}_{t_{ij}}$ and also the rigid body rotation around the contact point is accounted for to ensure that $ \B t_{ij}$ always remains in the local tangent plane of the contact \cite{01SEGHLP}.

   The translational and rotational acceleration of particles are calculated from Newton's second law; total forces and torques on particle $i$ are given by

      \begin{eqnarray}
\B F^{(tot)}_{i}&=& \sum_{j}\B F^{(n)}_{ij} + \B F^{(t)}_{ij}  \\
\B \tau ^{(tot)}_{i}&=& -\frac{1}{2}\sum_{j}\B r^{ij} \times \B F^{(t)}_{ij}.
\end{eqnarray}

   The tangential force varies linearly with the relative tangential displacement at the contact point as long as the tangential
   force does not exceed the limit set by the Coulomb law. When this limit is exceeded at the contact, we reset the magnitude of $t_{ij}$ so that $F^{(t)}_{ij} =\mu F^{(n)}_{ij}$ and this relationship holds until the particle loses contact. Thus the contact is considered to be stuck when $F^{(t)}_{ij} < \mu F^{(n)}_{ij}$ and is mobilized when the local yield criterion is satisfied.  A global damping is implemented to reach the static equilibrium in reasonable amount of time. After each compression step, a relaxation step is added so that the system reaches the static equilibrium. The global stress tensor is measured by taking averages of the dyadic products between the contact forces and the branch vector over all the contacts in a given volume,
  \begin{equation}
\sigma_{\alpha \beta} =\frac{1}{V}\sum_{j\neq i}\frac{r^{\alpha}_{ij} F^{\alpha}_{ij} }{2}
  \end{equation}
The pressure $P$ is determined from the trace of the stress. We cyclically compress and decompress the system in the quasistatic limit and after few cycles we do the measurements. In this work we focused on the decompression leg of the cycle i.e. approaching to the jamming point from above. Once the tangential force at a contact exceeds the coulomb threshold, it is considered to be a slip event and we calculate all the slip events that take place from the incremental reduction in pressure. All the calculations are done in the low pressure regime. As the pressure tends to zero, frictional packings will reach the generalized isostaticity line and then the scaling of the mechanical properties and the correlation length from this generalized isostatic point will be similar to the frictionless case.

Acknowledgements: This work was supported in part by the ISF-Singapore program and the US Israel Science Foundation.

\end{document}